\documentclass{article}

\usepackage{PRIMEarxiv}

\usepackage[utf8]{inputenc} 
\usepackage[T1]{fontenc}    
\usepackage{hyperref}       
\usepackage{url}            
\usepackage{booktabs}       
\usepackage{amsfonts}       
\usepackage{nicefrac}       
\usepackage{microtype}      
\usepackage{lipsum}
\usepackage{fancyhdr}       
\usepackage{graphicx}       
\graphicspath{{media/}}     
\usepackage{amsmath}
\usepackage{multicol}
\title{Video-rate quantitative phase and polarimetric imaging (QP²I) for label-free imaging of contracting muscle cells}

\author{
  Lucas Kreiss Roohian, Nachiket Pathak, David Ripp, Mitchell Harling, Nils M. Kronenberg, Marcel Schubert \\
  Humboldt Centre for Nano and Biophotonics,  \\
  University of Cologne \\
  Germany \\
   \And
  Malte C. Gather \\
  Humboldt Centre for Nano and Biophotonics,  \\
  University of Cologne \\
  Germany \\
  Cologne Excellence Cluster on Cellular Stress Responses in Aging-Associated Disease (CECAD),  \\
  University of Cologne \\
  Germany \\
  Centre of Biophotonics, School of Physics and Astronomy, \\
  University of St Andrews, \\
  UK \\
}

\begin{document}
\maketitle

\begin{abstract}
Quantitative phase imaging (QPI) has become a standard approach for label-free microscopy of transparent specimens, mapping the optical path length. However, the isotropic phase is blind to the molecular order underlying optical anisotropy. Implementations that also add quantitative contrast to birefringence rely on multi-frame illumination or polarization-state scanning, or on dedicated interferometric hardware and thus trade away acquisition speed or the simplicity of a standard microscope. We present Quantitative Phase and Polarimetric Imaging (QP²I), a computational imaging framework that converts a standard differential interference contrast (DIC) microscope with a polarization camera into a single-shot, video-rate, dual-contrast imaging system. We developed a new globally regularized phase reconstruction that eliminates artifacts of prior DIC-to-QPI approaches, providing quantitative phase information, and used per-pixel Stokes-based reconstruction to obtain quantitative anisotropic retardance. We validate the approach across simulations, static reference targets, fixed muscle tissue and living dynamically moving cells. QP²I enables all-optical and label-free studies of  contractile dynamics in live cardiomyocytes. It provides both structural and functional readouts at sufficient frame rates to resolve fast cardiomyocyte contractions. Conventional fluorescent labels bleach over time, perturb physiology, and preclude long-term monitoring. Instead, QP²I enables long-term, label-free interrogation of dynamic cellular processes at quantitative contrast and provides structural readouts similar to those provided by established fluorescent labels, such as frame-by-frame sarcomere length. It also offers previously inaccessible measurements, like quantitative analysis of retardance changes during muscle contraction, which provides a direct optical signature of muscle photoelasticity. 
\end{abstract}

\section{Introduction}

\begin{figure}[ht!]
\centering
\includegraphics[width=0.75\linewidth]{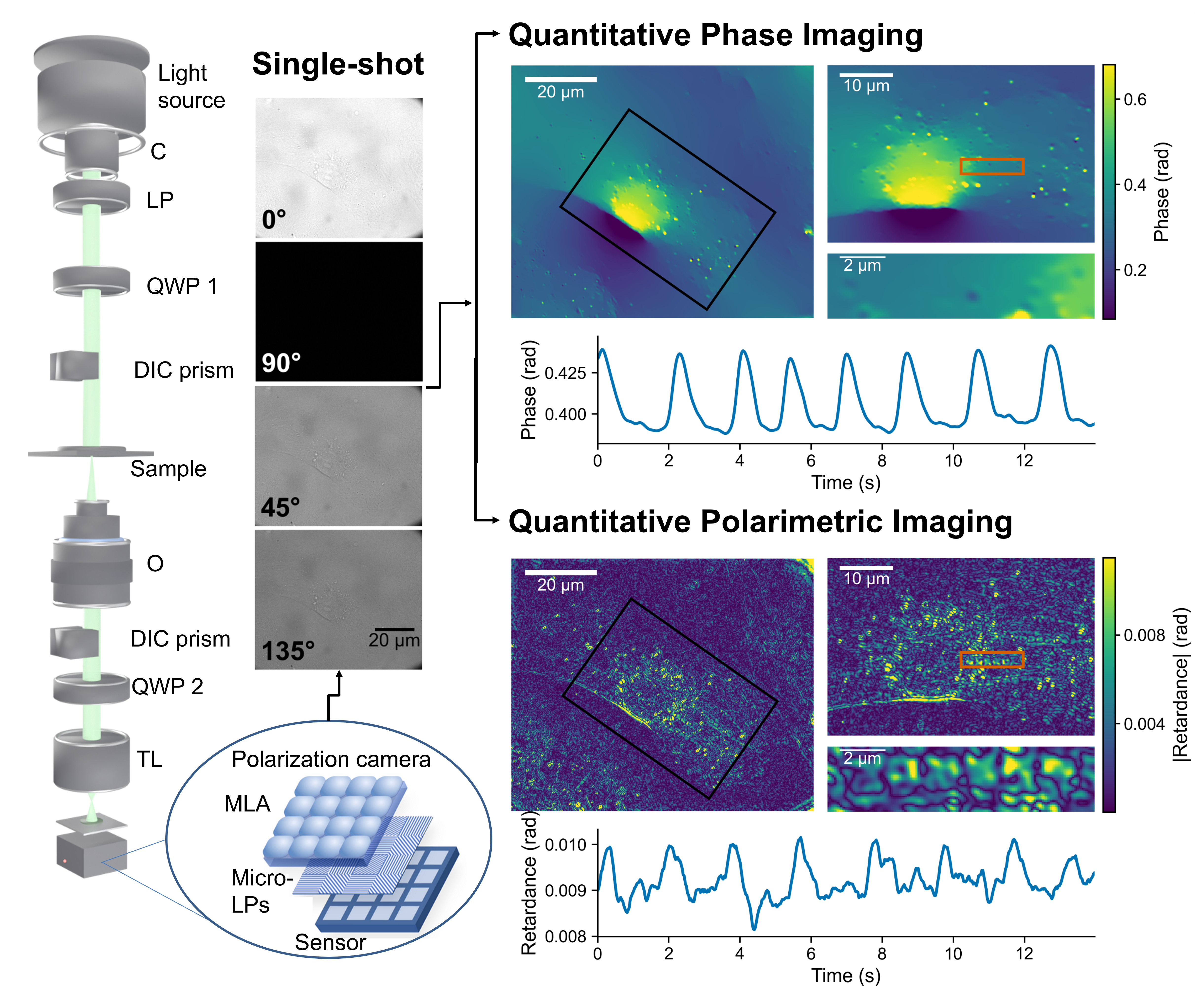}
\caption{\textbf{Quantitative phase and polarimetric imaging (QP²I) of the same, living cardiomyocyte.} Schematic of the full optical path. The raw data is composed of four polarization channels per acquisition ($I_{0}$, $I_{45}$, $I_{90}$, $I_{135}$). QP²I produced clean, artifact-free phase maps using the 'DIC-to-QPI' mode, with Nomarski prisms, as well as direct quantitative images of anisotropic retardance from the 'polarimetry' mode. While the former provides quantitative phase contrast based on refractive index and thickness of the cell, the latter gives quantitative access to single myofibrils. The single-shot acquisition of both modes provides these functional readout parameters at a speed that is only limited by the frame rate of the camera (70~fps). C = condenser, LP = linear polarizer, QWP = quarter waveplate, DIC = differential interference contrast, O = objective, TL = tube lens, MLA = micro lens array.}
\label{Fig1}
\end{figure} 

Quantitative phase imaging (QPI) is a widely used, fast and inexpensive label-free technique, that provides quantitative phase maps. Typically, computational phase retrieval is achieved by capturing several images at different conditions (e.g., different phase masks or illumination angles). For instance, QPI has been realized by acquiring four separate images with different phase masks as an add-on to conventional phase contrast microscopes~\cite{wang2011spatial}, or with different analyzer angles as add-on to conventional differential interference contrast (DIC) microscopes~\cite{nguyen2017gradient}. Fourier Ptychography microscopy (FPM) can also generate QPI maps using LED arrays for sequential angle-dependent illumination scanning~\cite{zheng2013wide,tian2014multiplexed}. More recently, polarization-dependent measurements and computational modeling of anisotropic phase properties have been implemented for FPM~\cite{xu2024tensorial, yeh2024permittivity, Yao2026}.

With the development of polarization sensitive cameras, single-shot, DIC-based QPI became possible by acquiring four different analyzer angles simultaneously, and computing quantitative phase gradient images~\cite{kwon2020single}. These can then be integrated in a phase map either via Hilbert-based methods~\cite{arnison2000using} or via line-artifact removal (LAR) after axis-aligned shear acquisition~\cite{yasuhiko2021single}. While effective for axis-aligned configurations, these line-by-line reconstructions remain susceptible to streak artifacts perpendicular to the DIC shear direction, which can corrupt the quantitative phase signal in regions of small gradient.

Single-shot polarimetry, in turn, has been demonstrated with polarization cameras to recover per-pixel Stokes parameters and molecular anisotropy at higher frame rates~\cite{bruggeman2024polcam}. Yet, to the best of our knowledge, no prior work has delivered both quantitative phase and quantitative polarimetry on the same instrument and from the same sample, or provided each from single frame acquisitions and at video rate.

Cardiomyocyte contractility is central to drug screening, disease modeling, and fundamental studies of muscle mechanics~\cite{gintant2019use,chaudhari2016identification,nguemo2012vitro}. In particular, iPSC-derived cardiomyocytes are now widely used as a platform for hypertrophic, dilated, and channelopathy-related disease models~\cite{eschenhagen2015modelling,smith2017human}. Functional imaging typically relies on fluorescent reporters, as calcium indicators, voltage-sensitive dyes, or fluorescently tagged sarcomere proteins. However, all fluorescent reporters eventually photobleach, while the high-power illumination in the UV or blue range exerts phototoxic stress, both of which severely limit long-term monitoring~\cite{icha2017phototoxicity}. Thus, there is a need for label-free, bio-compatible alternatives, capable of resolving contraction dynamics over long durations, at microscopic resolution, and at high frame rates.

Repeatedly capturing functional biological features during fast muscle contractions requires quantitative, time-resolved measurement of isotropic and anisotropic phase in single living cells, a measurement that has not been accessible at video rate. Existing approaches either lack quantitative polarimetric calibration~\cite{scalzo2021dense, lee2024piv}, operate too slowly to capture the contraction cycle~\cite{xu2024tensorial, yeh2024permittivity}, measure entire cell colonies instead of single cells~\cite{chaudhari2016identification,nguemo2012vitro} or rely on fluorescent labels that bleach and thus prevent extended recordings. Alternative label-free routes face different trade-offs. Electro-physiological impedance measurements do not offer single cell resolution and require expensive gold-coated well plates~\cite{chaudhari2016identification,nguemo2012vitro}. Tomographic and Fourier-ptychographic methods provide rich anisotropic and three-dimensional contrast~\cite{xu2024tensorial, yeh2024permittivity, yoon2025single}, but rely on tens to hundreds of illumination conditions, which precludes the investigation of fast cellular dynamics on the milisecond time scale. Second Harmonic Generation (SHG) microscopy~\cite{zipfel2003nonlinear} is particularly well-suited for 3D muscle tissue imaging~\cite{plotnikov2006characterization,both2004second,schneidereit2021advanced,schneidereit2024cas3d}; however, its point-by-point scanning approach typically prevents imaging at sufficient frame rates to resolve muscle contractions. Snapshot refractive-index tomography reaches video rate but at the cost of a strongly restricted field of view, since the camera sensor must be partitioned across many angular sub-images~\cite{wu2026snapshot}. Earlier multi-frame DIC approaches have combined phase and birefringence contrast~\cite{shribak2006orientation}, but require multiple sequential acquisitions at different specimen orientations and bias settings, again limiting their applicability to dynamic samples.

Here we present quantitative phase and polarimetric imaging (QP²I), a computational imaging framework that converts a standard DIC microscope equipped with a polarization camera into a single-shot, video-rate, dual-contrast functional imaging platform. QP²I is based on two complementary reconstructions that share the same instrument. To obtain quantitative phase contrast, we introduce a new single-stage global solver: a finite-difference operator matched to the DIC shear direction, combined with total-variation regularization, which recovers the phase map in one pass and inherently avoids the streak artifacts that have constrained earlier approaches. The reconstruction handles axis-aligned and diagonal shear configurations alike and requires only a single tunable parameter. Removing the Nomarski prisms converts the same instrument to a polarization microscope, enabling a per-pixel Stokes analysis that yields quantitative data on retardance and molecular orientation. Together, the two modes provide complementary contrast on the same biological structures and across co-aligned fields-of-view. The isotropic phase is linked to local mass density and isotropic optical path length, while polarimetry isolates anisotropic, optically ordered components such as the contractile apparatus of muscle cells.
We validate QP²I across simulations, static reference targets, and biological samples. Finally, we demonstrate sustained, label-free frame-by-frame readouts of sarcomere length, and quantitative time-resolved retardance during the contraction of individual myofibrils in cardiomyocytes, providing a direct optical window into the photoelastic response of the heart.

\section{Principle of QP²I}

\subsection{Optical setup}
The optical setup consisted of a conventional DIC microscope (TiE2, Nikon) equipped with a polarization sensitive camera (BFS-U3-51S5P-C USB 3.1 Blackfly, Sony IMX250MZR, 2,448 x 2,048 pixels at 3.45~x~3.45~µm², 70~fps) and a quarter waveplate (QWP~2, see Fig.~\ref{Fig1}). In the illumination path, a combination of rotatable linear polarizer and quarter waveplate act as de Sénarmont compensator, that produces polarized light of tunable ellipticity based on the LP rotation angle. The linear polarizer was rotated to create circularly polarized light in the illumination path. Additionally, we used the built-in green filter of the microscope to ensure illumination across a narrow wavelength band and hence avoid wavelength-dependent dispersion effects. For the QPI mode, two respective Nomarski prisms were inserted, and the amount and direction of shear were determined by imaging sub-resolution particles. For the polarimetric imaging mode, the Nomarski prisms were removed. In each case, the four polarization channels, recorded simultaneously by the camera, were split and fed into the respective reconstruction, as described below. Therefore, each 2~x~2 grid of micro-polarizers of the polarization camera is treated as one virtual super pixel that carries all four polarization channels. Thus, the effective resolution of the reconstructions is 1,224 x 1,024 pixels. Fluorescence images of labeled cells were captured under blue epi-illumination (460~nm) using a GFP filter cube (3035D, 472/30 excitation and 520/35 emission) and a separate camera (Hamamatsu ORCA flash 4.0 C13440). All samples, except fixed muscle fibers (Fig.~\ref{Fig3}i-l), were imaged using a 100x oil immersion objective (Nikon Plan Apo 100x/1.4 DIC N2). Fixed muscle fibers were imaged using a 40x air objective (Nikon Plan Fluor 40x/0.75 DIC N2).

\subsection{Single shot quantitative phase imaging}
The polarization camera enables simultaneous, single-shot acquisition of four different DIC images with different analyzer angles. For quantitative phase reconstruction, these four polarization channels are used to extract the quantitative phase gradient image (QPGI) via the 2-argument arctangent:
\begin{equation}
  \phi_{\text{QPGI}} = \arctan2(I_{45}-I_{135} , I_0-I_{90})
\end{equation}

where $I_{p}$ is the intensity image of all pixels at the respective polarizer direction $p$ in degrees. The circularly polarized illumination introduced a polarization bias $\eta_{\text{bias}}$ in the background that was estimated empirically from background pixels, and subtracted from the QPGI. Finally, the QPGI was scaled by the calibrated shear distance and integrated to obtain the final phase map. Previous works on single-shot QPI already exploited this relation~\cite{kwon2020single,arnison2000using,yasuhiko2021single}, for instance by integrating this gradient via Hilbert-based spectral methods~\cite{arnison2000using} or cumulative summation followed by line-artifact removal (LAR) as proposed by Yasuhiko et al.~\cite{yasuhiko2021single}. Our method was tested against the latter, since it was shown to produce high-quality phase images, but requires elaborate fine tuning of up to four essential hyper parameters (the regularization parameter for high-pass filter $\mu$, the regularization parameter for background $\xi$, the high-pass and low-pass filters for shear artifact removal). Moreover, LAR requires a reliable background mask that directly enters the reconstruction itself (see Table~\ref{tab_simulation}). Its background term drives the reconstructed phase to zero on every pixel the mask labels as background. Outside the cell, the phase is therefore set by the mask rather than by the measurement, and the reconstruction is only as reliable as the segmentation. To avoid this and improve overall robustness, we formulate a new phase recovery as a single-stage variational problem with a single hyper parameter for total variation (TV) regularization:

\begin{equation}
  \min_\phi \tfrac{1}{2}\|D_{\text{shear}}\phi - \mathrm{QPGI}\|^2 + \lambda_{TV} \cdot \mathrm{TV}(\phi)
  \label{DCT_TV}
\end{equation}

with $D_{shear}$ as a continuous directional difference operator matched to the microscope's physical shear vector and $\lambda_{TV}$ as the weighting factor for TV regularization. Our reconstruction aims to minimize the difference between the measured QPGI and the shear difference of the recovered phase $\phi$, while simultaneously minimizing TV. This optimization problem \eqref{DCT_TV} is solved using the alternating direction method of multipliers (ADMM), wherein the quadratic $\phi$-subproblem is solved via a preconditioned conjugate gradient (PCG) method utilizing a Discrete Cosine Transform (DCT)-based Poisson solver as a structural preconditioner under Neumann boundary conditions. In simulation, different TV weights $\lambda_{TV}$ were tested to obtain the optimal result. Finally, the background phase was subtracted from the full image, to pull the background phase of the surrounding medium to 0. The background mask therefore enters only before and after the solve, for the bias estimate, the global sign and the zero offset, and for computing quality metrics, but never as a constraint on $\phi$, as it is done in LAR. For video-rate processing, each new frame was initiated with the solution from the previous frame to reduce per-frame computation. 

\subsection{Single shot quantitative polarimetric imaging}
For polarimetric imaging, the Nomarski prisms were removed and the same polarization camera was used to analyze the polarization state of light transmitted through the sample. Here, the circularly polarized illumination light directly shines on the sample, where birefringent structures introduce a differential phase shift (retardance) between its fast and slow axes, converting the initially circular polarization into an elliptical state. The degree of ellipticity encodes the retardance $\eta$, while the orientation of the ellipse encodes the fast-axis angle $\theta$. The Stokes parameters at the detector $S_{d,i}$ can be computed directly from the four polarization channels of the camera and then normalized to $S_{d,0} = I_0 + I_{90}$ as:
\begin{multicols}{2}
  \begin{equation}
    S_{d,1} = \frac{(I_0 - I_{90})}{I_0 + I_{90}} 
  \end{equation}\break
  \begin{equation}
    S_{d,2} = \frac{(I_{45} - I_{135})}{I_0 + I_{90}} 
  \end{equation}
\end{multicols}
Crucially, the camera pixels with four linear micropolarizers are inherently insensitive to circular polarization and thus, $S_{d,3}$ is unknown (note that within the Stokes formalism, unknown does not mean $S_{d,3}=0$). However, this Stokes parameter is essential to solve the sample's retardance $\eta$. This issue was resolved by the second quarter waveplate $QWP_{d}$ that was already placed before the camera. It converts the circular component of the light into a linear component that the camera can measure. Effectively, the QWP rotates the Stokes vector such that the sample's circular polarization information ($S_{s,3}$, where sub-script $_s$ denotes sample) is encoded in the measurable linear channels detected by the polarization camera that is $S_{d,0}, S_{d,1}$ and $S_{d,2}$ (see Supplementary Information for more details). 

To then recover the sample's original Stokes vector $S_s=[S_{s,0}, S_{s,1}, S_{s,2}, S_{s,3}]$, we express the measured Stokes parameters $S_d$ as a combination of the sample properties and $QWP_{d}$:
\begin{equation} 
S_{d} = M_{QWP} \cdot S_s
\label{eq:mueller}
\end{equation}

Inverting Eq.~\eqref{eq:mueller} then yields the target Stokes components of the sample:
\begin{equation}
S_{s} = M_{QWP}^{-1} \cdot S_{d} = ( 1,\, S_{d,1},\, -S_{d,3},\, S_{d,2})^T 
\label{eq:mueller_inv}
\end{equation}

Thus, the measured retardance $\eta_m$ can be obtained as:
\begin{equation}
\eta_m = \arccos(S_{s,3}) = \arccos(S_{d,2}) 
\label{eq:eta_raw}
\end{equation}

However, this term still contains the circular polarized illumination component $\eta_{\text{bias}}$ (see above), which can be estimated as empirical bias from the background pixels and subtracted to obtain the retardance of the sample $\eta$ as: 
\begin{equation}
\eta =\left(\arccos(S_{d,2}) - \eta_{\text{bias}}\right)
\label{eq:eta}
\end{equation}

While the addition of the QWP at this angle allows to accurately deduce the sample's retardance from the detectable Stokes parameters $S_{d,1}$ and $S_{d,2}$, the Stokes parameter $S_{s,2}$ is now expressed by the entirely unknown $S_{d,3}$, which would be required to solve for the sample's fast-axis orientation $\theta$:
\begin{equation}
\theta = \tfrac{1}{2} \operatorname{atan2}(-S_{s,1},\; S_{s,2}) = \tfrac{1}{2} \operatorname{atan2}(-S_{d,1},\; -S_{d,3})
\label{eq:theta}
\end{equation}

In the present work, we solve this by exploiting the degree-of-polarization (DOP):
\begin{equation}
DOP = \frac{\sqrt{S_{d,1}^2 + S_{d,2}^2 + S_{d,3}^2}}{S_{d,0}}
\end{equation}

Assuming fully polarized light after the sample (DOP = 1), the unknown $S_{d,3}$ can be expressed as:
\begin{equation}
|S_{d,3}| = \sqrt{S_{d,0}^2 - S_{d,1}^2 - S_{d,2}^2}
\label{eq:dop}
\end{equation}

to solve Eq.~\ref{eq:theta}. The assumption of fully polarized light is reasonably valid for the relatively thin, weakly-scattering samples used in this study, where depolarization from multiple scattering is negligible. However, while $\eta$ can be solved analytically and follows directly from the well determined equation Eq.~\ref{eq:eta}, the solution for $\theta$ is not as robust for all sample types (see Supplementary Information). According to the directionality of the phase shift with respect to the illumination bias, the reconstructed retardance is signed positive (same direction as bias) or negative (opposite to bias). In this work, however, the retardance is visualized as absolute value $|\eta|$. Alternatively, examples of signed retardance maps are shown in Supplementary Fig.~S3. 

\section{Results}

\begin{figure}[ht!]
\centering
\includegraphics[width=0.5\linewidth]{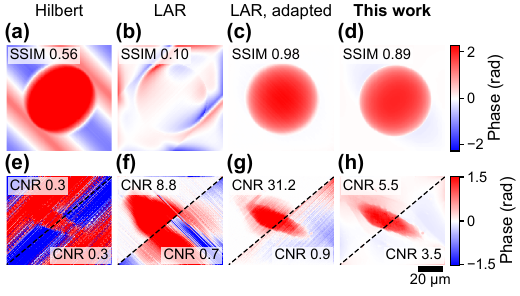}
\caption{\textbf{Validation of the phase reconstruction of QP²I on simulated and real cell data.} (a)--(d) Reconstructions for simulated input with 1.07\% Gaussian noise added for (a) Hilbert integration, (b) LAR as originally published~\cite{yasuhiko2021single}, (c) adapted LAR for diagonal shear and (d) this work. Each case states the structure similarity index metric (SSIM) between simulated ground truth and reconstruction. (e-h) The same four phase reconstruction methods applied to data of a live cardiomyocyte. In each case, the solver was tested with known manually segmented background (see the half above the dashed diagonal line) and with unknown background (see the half below the dashed diagonal line). The contrast-to-noise ratio (CNR) on that known background is given top left, the CNR on the unknown background bottom right. Detailed metrics in Tables~\ref{tab_simulation} and ~\ref{tab_live_cell}. }
\label{Fig2}
\end{figure}

\begin{figure}[ht!]
\centering
\includegraphics[width=0.5\linewidth]{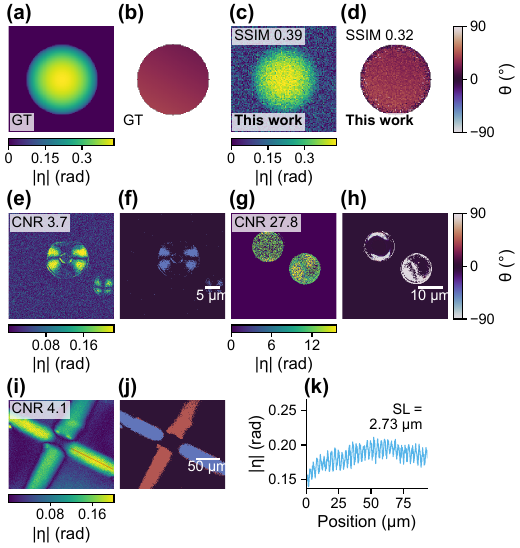}
\caption{\textbf{Validation of the polarimetric reconstruction of QP²I on simulated and real data.} (a)--(d) Optical simulations with ground-truth (a\&b) and reconstructions of retardance (c) and orientation (d) using simulated input with 1.07\% Gaussian noise added. (e-l) The same polarimetric reconstructions on real samples: (e\&f) a starch grain; (g\&h) LCE particles; and (i-l) rat muscle fibers. (k) line profile of the retardance along one of the muscle fibers, modulated by the sarcomere pattern with a sarcomere length of 2.73\,$\mu$m (see supplementary information for more detail).}
\label{Fig3}
\end{figure} 

\begin{table*}[]
  \caption{Performance of the phase reconstruction methods on simulated data with 1.07\% added noise and 2 averaged frames. Root mean squarred error (RMSE) and structure similarity index (SSIM) are computed between reconstruction and known ground truth. 'Mask in solver' states if the background mask is a constraint inside the reconstruction itself. The simulation supplies an exact mask, the most favourable case for LAR; DCT+TV uses the mask only to remove the constant bias and tilt of the
  QPGI and to set the global sign before the solve. Parameters used: LAR and adapted LAR $\mu = 500$, $\xi = 50$; DCT+TV $\lambda_\mathrm{TV} = 0.05$. $f_\mathrm{lo}$, $f_\mathrm{hi}$: spectral cut-offs of LAR. Best values in each column are \textbf{bold}.}
  \centering
  \begin{tabular}{|l|l|l|c|c|c|c|}
    \hline
    Method & Tunable & Mask in & \multicolumn{2}{c|}{Isotropic phase imaging} & \multicolumn{2}{c|}{Anisotropic polarimetric imaging} \\
     & parameters & solver & RMSE ($\downarrow$) & SSIM ($\uparrow$) & RMSE ($\downarrow$) & SSIM ($\uparrow$) \\
    \hline
    Hilbert integration of QPGI~\cite{arnison2000using} & \textbf{0} & \textbf{No} & 1.970 & 0.564 & N/A & N/A \\
    \hline
    LAR integration of QPGI~\cite{yasuhiko2021single}, & 4: $\mu, \xi, f_\mathrm{lo}, f_\mathrm{hi}$ & Yes & 1.618 & 0.096 & N/A & N/A \\
    \quad as published (axis-aligned shear) & & & & & & \\
    \hline
    LAR integration of QPGI, adapted & 4: $\mu, \xi, f_\mathrm{lo}, f_\mathrm{hi}$ & Yes & \textbf{0.091} & \textbf{0.982} & N/A & N/A \\  
    \quad to diagonal shear (this work) & & & & & & \\
    \hline
    This work (DCT+TV) & 1: $\lambda_\mathrm{TV}$ & \textbf{No} & 	0.298 & 0.894 & \textbf{0.060} & \textbf{0.391} \\
    \hline
  \end{tabular}
  \label{tab_simulation}
\end{table*}

\subsection{Validation and benchmarking of quantitative phase imaging}
The performance of the QPI modality was first evaluated on simulated image data of 128~$\times$~128 pixels (assuming 0.115 $\mu m$ pixel size) with dielectric spheres in the center (diameter: 10~$\mu m$) embedded in a homogeneous medium (n=1.48), with Gaussian noise added at varying levels. For isotropic phase testing (Fig.~\ref{Fig2}a-d), the simulated sphere had a refractive index of n=1.5. The illumination was simulated as monochromatic light at 550~nm with a bias retardance of $\pi/2$ rad. In this simulation, we compared our phase retrieval method against Hilbert-based integration~\cite{arnison2000using}, and a cumulative-sum integration with line-artifact removal (LAR)~\cite{yasuhiko2021single}; all applied to the same simulated DIC data with diagonal ($-45^{\circ}$) shear. Because LAR was originally formulated for shear along an image axis~\cite{yasuhiko2021single}, we additionally implemented a version adapted to diagonal shear: the QPGI image is rotated by $45^{\circ}$ so that the shear lies along $x$, reconstructed with the $x$-shear LAR solver, and the resulting phase is rotated back. The regularization parameters ($\lambda_{TV}$ for DCT+TV; $\mu$ and $\xi$ for LAR, with its two spectral cut-offs at their default values) were swept over reasonable ranges and the best combination was used. At a representative noise level of 1.07\% (Fig.~\ref{Fig2}a-d, Fig.~\ref{Fig3}b\&d, Table~\ref{tab_simulation}), the Hilbert reconstruction showed the characteristic overestimation and directional streak artifacts perpendicular to the shear direction, and LAR as published left residual diagonal streaks, consistent with the known limitations of line-by-line integration for non-axis-aligned shear. Our new phase retrieval recovered a smooth phase map with the correct spatial profile, with a mild amplitude underestimation caused by the TV regularization. In this simulation the adapted LAR reached the lowest error (RMSE 0.091; SSIM 0.982), however, at the cost of four tunable parameters. Moreover, LAR requires a reliable background mask that directly enters the reconstruction itself as a constraint. In simulation, this mask is available and perfect. However, in live-cell imaging a perfect background mask is difficult to obtain, especially for images of multiple cells or moving cells. 

Although segmentation models such as SAM~\cite{kirillov2023segment} (see Supplementary Material) can provide reasonably accurate masks, phase solvers that require perfect masks to push the background phase to zero essentially turn phase retrieval into a problem of cell segmentation. Instead, our new DCT+TV solver is agnostic of background and foreground and has only a single tunable parameter. Here, background pixels are only used to estimate the bias retardance and the noise level to compute SNR or CNR. To test how much each method depends on its background mask for a real live cell measurement, we reconstructed a single frame of a live cardiomyocyte recording with and without the background constraint. Fig.~\ref{Fig2}e-h then shows the reconstructions with known background in the top-left half and the reconstruction of the same image without known background on the bottom-right half. Because LAR forces its solution to zero on every background pixel it is given, the adapted LAR reached a CNR of 31.2 on the known half, but drops 35-fold to only 0.9 on the held-out half. Instead, our new phase retrieval (5.5 and 3.5) gave similar values on both halves, since neither uses the mask in its solver. Hilbert integration (0.3 and 0.3) and LAR as published~\cite{yasuhiko2021single} did not recover the cell accurately at this diagonal shear. The high CNR of LAR on known background is therefore inflated by forcing the phase to zero outside the mask rather than relying on the measurement. Moreover, this makes the LAR method much more susceptible to uncertainty in the segmentation mask as well as to measurement noise. In a simulated noise series between 0-15\% added noise (supplementary Fig.~S2), the SSIM of LAR and adapted LAR drop significantly with increasing noise, while our approach stays unaffected and even outperforms LAR with perfectly known background mask for noise levels >12\%. 

\begin{table}[]
  \caption{Contrast-to-noise ratio of the phase reconstructions of a live cardiomyocyte (raw frame 0 of the QPI recording). Parameters: LAR $(\mu, \xi) = (500, 50)$ and adapted LAR $(500, 10)$, optimal in the simulation (Table~\ref{tab_simulation}); $\lambda_\mathrm{TV} = 0.05$, as in the video reconstruction. CNR $= (\langle\phi_\mathrm{cell}\rangle - \langle\phi_\mathrm{bg}\rangle) / \sigma(\phi_\mathrm{bg})$, negative when the phase of the cell comes out inverted. \emph{Background known}: the solver is given the complete background mask. \emph{Background unknown}: the same frame without a mask; where a solver needs one (offset and tilt of the phase gradient, the phase zero) the whole frame takes its place, and the background term of both LARs is switched off ($\xi = 0$, $\mu$ as above). The sign of each reconstruction without a mask is taken from its counterpart with the mask. Both CNRs are taken over the whole frame with the complete mask. Best values in each column are \textbf{bold}.}
  \centering
  \begin{tabular}{|l|c|c|}
    \hline
     & \multicolumn{2}{c|}{|CNR| ($\uparrow$)} \\
     & known & unknown  \\
    Method & background & background \\
    \hline
    Hilbert~\cite{arnison2000using} & 0.31 & 0.30 \\
    \hline
    LAR~\cite{yasuhiko2021single}, as published & 8.75 & 0.72 \\
    \hline
    LAR, diagonal shear & \textbf{31.21} & 0.90 \\
    \hline
    This work & 5.45 & \textbf{3.48} \\
    \hline
  \end{tabular}
  \label{tab_live_cell}
\end{table}

\subsection{Validation and benchmarking of quantitative polarimetric imaging}
As before, we started by evaluating the polarimetric imaging in optical simulations. Therefore, a sphere of anisotropic retardance and angular orientation of its fast axis was simulated with 0.005 rad of retardance per pixel of thickness. At 0.115 µm/px and 550 nm that is $\Delta n \approx 3.8 × 10^{-3}$ (Fig.~\ref{Fig3}a-d). The analytical Stokes-based reconstruction recovered retardance with a SSIM of 0.39 at 1.07\% noise level. Since this Stokes-based reconstruction is an exact analytical solution, the solutions for retardance and orientation at zero-noise are perfect (supplementary Fig.~S2b, SSIM = 1.0). However, the noise sweep reveals that both retardance and orientation SSIM dropped sharply at the onset of noise ($\sim$2\%) and remained approximately constant at higher noise levels ($>$5\%), indicating that the reconstruction quality is limited by the inherent sensitivity of the Stokes parameter inversion rather than by the absolute noise magnitude. Because the retardance is recovered through an inverse cosine, zero-mean intensity noise is mapped onto strictly non-negative retardance values, so the background, whose true retardance is zero, acquires a small positive noise floor (Fig.~\ref{Fig3}c,e,i). This floor barely affects the RMSE, but SSIM normalizes the comparison by the local signal variance and is therefore dominated by the large, structure-free background, where any deviation counts fully. The retardance reconstruction was notably more robust than orientation across all noise levels, consistent with the fact that retardance depends on the magnitude of the Stokes vector components whereas orientation depends on their ratio (see Table~\ref{tab_simulation} and Supplementary Information).

Next, we tested the polarimetric imaging mode of QP²I in a range of samples with different optical characteristics, including corn starch grains, highly birefringent LCE beads and fixed muscle fibers (see supplementary Methods). Fig.~\ref{Fig3}e\&f show the polarimetric reconstruction of two corn starch grains immersed in oil. The retardance map revealed the characteristic thickness-dependent birefringence profile of starch granules, with peak retardance values of approximately 0.16~rad. Assuming close to rotational symmetry for grains in a range of $6-13~\mu m$ diameter, the effective birefringence would be within $\Delta n \approx 1 - 2.1 \times 10^{-3}$, which is consistent with previously reported values for corn starch~\cite{litvyak2018size}. Additionally, the orientation map (Fig.~\ref{Fig3}f) resolved the well-known radial fast-axis pattern of starch granules, with the reconstructed angles varying smoothly across each grain and the two grains exhibiting mutually consistent orientation fields despite differing spatial positions and orientations. 

To validate our retardance unwrapping framework on strongly birefringent specimens, we imaged custom LCE microspheres as shown in Fig.~\ref{Fig3}g\&h. The reconstructed retardance map (Fig.~\ref{Fig3}g) demonstrate successful recovery of retardances far exceeding the principal $[0,\pi]~\text{rad}$ interval, reaching unwrapped values above $10~\text{rad}$. Crucially, the fast-axis orientation map (Fig.~\ref{Fig3}h) reveal a uniform orientation field within the particle interior ($\sim -36^\circ$) that is completely free of the $90^\circ$ jump artifacts typically induced at half-wave wrap boundaries. This confirms that wave-order determination successfully restores both quantitative retardance magnitudes and physically continuous fast-axis orientations. 

To demonstrate the applicability to biological tissue, crossed skeletal muscle fibers from rats were imaged in polarimetric mode (Fig.~\ref{Fig3}i--l). The retardance map clearly resolved the two fiber bundles with distinct retardance values, and the orientation map unambiguously separated their crossing directions, appearing as opposing signs in Supplementary Fig.~S3 and their orientation angles are shifted by 90$^\circ$ in Fig.~\ref{Fig3}j. A line profile extracted along a single fiber (red ROI of Fig.~\ref{Fig3}i\&j) revealed periodic retardance oscillations with an amplitude of approximately 0.02~rad superimposed on a baseline of $\sim$0.18~rad, corresponding to the alternating birefringent A-bands and less birefringent I-bands of the sarcomere structure. The measured retardance of 0.15--0.20~rad is consistent with the expected order of magnitude for muscle tissue (assuming a birefringence of $\Delta n \approx 1.79 \times 10^{-3}$~\cite{mcbride2022vivo}). The corresponding FFT of the retardance line profile (supplementary Fig.~S1) yielded a single, well-defined peak indicative of a sarcomere length of 2.73~$\mu m$. 

\subsection{Live cell imaging of contracting CDMs}
\begin{figure}[ht!]
    \centering
    \includegraphics[width=0.6\columnwidth]{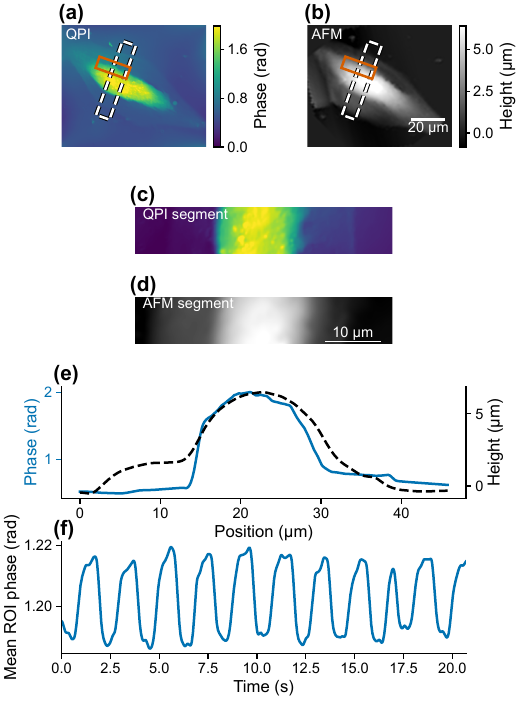}
    \caption{\textbf{Quantitative phase results of a live cardiomyocyte.} (a)~Phase reconstruction showing the dominant phase contribution from the cell nucleus, next to (b)~a co-registered atomic force microscopy (AFM) height map of the same cell. (c)~\&~(d)~Zoomed in views of QPI and AFM respectively. (e)~Line profiles across the cell center comparing the reconstructed phase and AFM height, showing high spatial agreement with a peak height of $\sim6\mu m$ corresponding to $\sim$2~rad phase. (f)~Time-resolved phase dynamics of a live, spontaneously contracting cardiomyocyte acquired at 33~fps. The mean phase within the zoomed region of interest is plotted over time, with contraction events producing a periodic oscillation. }
    \label{Fig4}
\end{figure}

We further applied QP²I to living, dynamically moving biological samples. Both QPI and polarimetric imaging modes were applied to the same cells by inserting or removing the Nomarski prisms between acquisitions. Fig.~\ref{Fig4} shows our phase reconstruction and an AFM topography map of the same cell. Line profiles across the cell (Fig.~\ref{Fig4}e) show a high level of agreement between the cell thickness measured by AFM and the phase profile. Taking the AFM height of the same cell as its thickness d, the phase above the background along the profile in Fig.~\ref{Fig4}e (0.60 – 0.77~rad over d = 3.2 – 4.6~µm) corresponds to an average refractive index of $n_{cell} = n_{medium} + \Delta \phi \lambda/(2\pi d)$ of 1.347 – 1.350 (assuming $n_{medium} = 1.334, \lambda = 550 nm$). This is perfectly within the range of live cardiomyocytes as reported previously by intracellular microlasers~\cite{schubert2020monitoring}. While AFM provides absolute height information at nanometer axial resolution, it is fundamentally limited in temporal throughput (scan rate of 16~pixels per second, resulting in a frame rate of 0.0038~fps, see supplementary material), and requires physical contact with the sample, rendering it incompatible with live-cell dynamics. Our single-shot QPI, by contrast, operates at 87.7 megapixels/s or 70~fps (1,224~$\times$~1,024 pixels per reconstruction, 2,448~$\times$~2,048 pixels per raw acquisition), enabling continuous monitoring of cellular morphology changes. Fig.~\ref{Fig4}~f demonstrates this capability by showing the time-resolved phase dynamics of a spontaneously contracting cardiomyocyte, where sub-second events are captured without shear-induced spatial or motion-induced temporal artifacts. Notably, our phase reconstruction remains artifact-free even during contraction, where steep local phase gradients from rapid cell deformation might produce streak artifacts in conventional Hilbert or LAR-based methods. The 20s QPI trace in Fig.~\ref{Fig4}f shows 11 full contractions, corresponding to a contraction frequency of 0.55~Hz.  During the contraction cycle, the mean phase in the ROI of Fig.~\ref{Fig4}f changes by $\sim$5\%. Under the simplifying assumption of constant thickness, which does not strictly hold because the cell compresses and relaxes, this would correspond to a refractive index modulation between 1.3487 and 1.3495, which is again of the same order as the oscillation between 1.348 - 1.354 that has been reported during CDM contractions by intracellular microlaser and were attributed to changes in sarcomeric protein density~\cite{schubert2020monitoring}.

However, while QPI captures the overall cell morphology and the temporal dynamics of thickness and refractive index with high sensitivity, it fundamentally cannot visualize the contractile machinery that causes these contractions. The isotropic phase maps of live cardiomyocytes are dominated by the cell nucleus, whose elevated refractive index and substantial thickness produce a pronounced phase peak that overshadows the surrounding cytoplasm (Fig.~\ref{Fig4}). The contractile apparatus of the sarcomeric myofibril network, which is responsible for the mechanical contraction, contributes comparatively little to the isotropic phase signal, as the refractive index contrast between myofibrils and the surrounding cytoplasm is minimal. 

\begin{figure}[ht!]
    \centering
    \includegraphics[width=0.4\columnwidth]{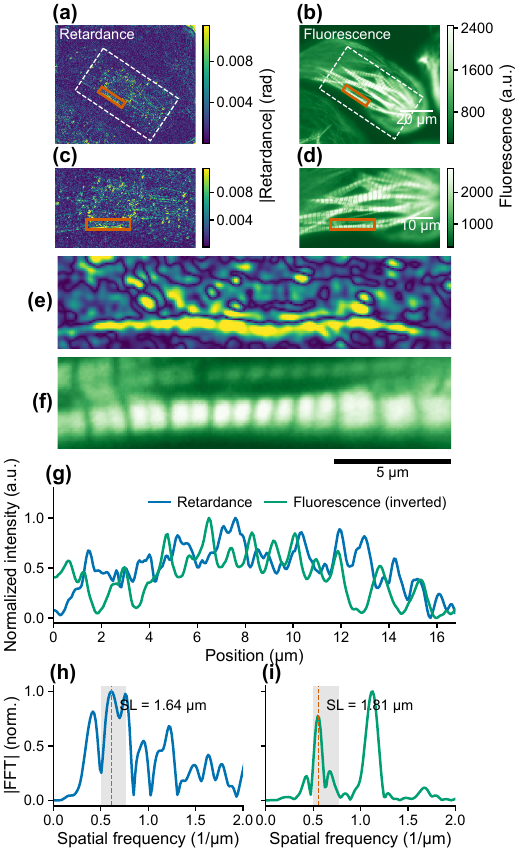}
    \caption{\textbf{Quantitative polarimetric imaging of a live cardiomyocyte.} (a)~Full field-of-view reconstructions of retardance, and (b)~co-registered fluorescence, with zoomed in views (c~\&~d) and an individual myofibril (e~\&~f). The retardance map reveals the myofibrillar network at milli-radian sensitivity. (g)~Line profiles along the myofibril axis for normalized retardance and fluorescence, showing periodic modulations corresponding to the alternating A-bands and I-bands of the sarcomere structure. The retardance profile exhibits higher band-to-band contrast and resolves finer sub-sarcomeric features not captured by the fluorescence signal. (h~\&~i)~Normalized FFT spectra, with peaks indicating sarcomere lengths.}
    \label{Fig5}
\end{figure}

On the other hand, the birefringence of the aligned myosin and actin filaments within sarcomeres produces a measurable anisotropic retardance signal. The secondary polarimetric imaging contrast provides direct access to the structural organization and orientation of the contractile apparatus. Live iPSC-derived CDMs were imaged followed by fluorescence imaging of labeled sarcomeres (Fig.~\ref{Fig5}). Both revealed the myofibrillar network beneath and around the nucleus that could not be observed in conventional phase contrast. The retardance map resolved individual myofibrils, corresponding to an effective birefringence sensitivity of $\Delta n_{\text{eff}} \approx 2.2-4.4 \times 10^{-4}$ for an estimated myofibril thickness of $\sim$1--2~$\mu$m. It was also possible to image a CDM with two adjacent myofibrils that were oriented nearly perpendicularly (supplementary Fig.~S6), and the measured signed retardance showed a sign flip between both myofibrils orientated at 90$^{\circ}$ difference. This quantitative angular discrimination is analogous to the crossed-fiber result obtained for bulk muscle tissue (Fig.~\ref{Fig3}i-l), but here achieved on the single cell level where total retardance is significantly lower. 

The co-registered fluorescence confirms the spatial correspondence of the birefringent structures with the labeled sarcomeric proteins, validating that the polarimetric contrast originates from the organized contractile apparatus rather than other cellular components. Zooming onto an isolated myofibril (Fig.~\ref{Fig5}e\&f, supplementary Fig.~S5e\&f and S6e\&f) in our retardance reconstruction revealed the repeating pattern of sarcomeres, comparable to that shown by fluorescence reference. The retardance line profile exhibits periodic oscillations corresponding to the alternating birefringent A-bands (myosin-rich) and less birefringent I-bands (actin-dominated), with notably higher modulation contrast between these bands than observed in the fluorescence channel. Furthermore, the retardance profile resolves fine sub-sarcomeric features at better contrast than in the fluorescence images of the same myofibril. For any myofibril we tested, an FFT-based analysis yielded a sarcomere length (SL) of our label-free retardance measurement that agrees with the fluorescence derrived SL within the optical resolution limit of our system (see supplementary information), confirming that the polarimetric reconstruction preserves the spatial periodicity of the sarcomeric lattice with sufficient quantitative accuracy for contractility analysis. The FFT analysis shows higher harmonics at 2~$\times$, 4~$\times$ (Fig.~\ref{Fig5}i). However, the bandpass filter (gray band) for determining the sarcomere length successfully rejected these. 

\subsection{Video-rate birefringence imaging of beating cardiomyocytes}

\begin{figure}[ht!]
\centering
\includegraphics[width=0.45\linewidth]{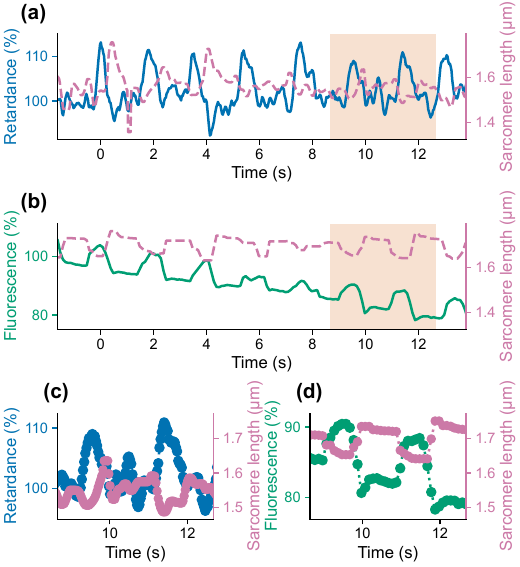} 
\caption{\textbf{Time-resolved sarcomere dynamics of live cardiomyocytes.} (a)~Top-5\% quantile of retardance (left axis) and retardance-derived sarcomere length (right axis) within a single myofibril ROI, acquired at 66 fps, averaged pairwise to 33 fps. (b)~Corresponding mean fluorescence intensity (left axis) and fluorescence-derived sarcomere length (right axis) from the co-registered video, acquired at 10~fps. (c)~Zoomed view of a single contraction cycle from the retardance channel, and (d)~in the fluorescence channel. }
\label{Fig6}
\end{figure} 

A central motivation for label-free contractility imaging is the ability to sustain high-speed recordings without the temporal and phototoxic constraints imposed by long-term fluorescence microscopy. Next, we therefore compared the ability of polarimetric imaging and conventional fluorescence imaging for the tracking of sarcomere length over time. We first recorded polarimetric retardance data at an effective frame rate of 33~fps, i.e., averaging pairs of subsequent frames to reduce reconstruction noise. The resulting evolution of retardance and sarcomere length calculated from it clearly show individual contraction–relaxation cycles (Fig.~\ref{Fig6}a). Subsequently, a fluorescence reference was acquired from the same cardiomyocyte at 10~fps acquisition rate (i.e., 3$\times$ slower sampling than the retardance imaging). Fig.~\ref{Fig6}b shows the resulting evolution of fluorescence intensity and of the sarcomere length calculated from it, with the first contraction co-registered to the first contraction obtained from retardance-imaging. Both modalities captured 8 full contractions over the 13-second-long time-traces, corresponding to a contraction frequency of 0.66~Hz, which is comparable to the 0.55~Hz frequency observed earlier from the QPI video of a different cell (Fig.~\ref{Fig4}f). In both modalities, signal intensity increases when sarcomere length decreases, and vice versa. This consistent trend further corroborates that both methods capture the same underlying contractile dynamics. 

A zoomed in comparison of two consecutive contraction events (Fig.~\ref{Fig6}c,d) shows that the faster acquisition speed of the retardance imaging resolves additional contraction features that are not visible in the fluorescence trace. Specifically, the retardance trace resolves fine temporal dynamics, within a single contraction event, with the contraction waveform exhibiting an intermediate relaxation and secondary peak before the final relaxation phase. These features are smoothed into a featureless plateau in the fluorescence channel (Fig.~\ref{Fig6}d), which at 10~fps has a Nyquist-limited temporal resolution of $\sim$200~ms or $\sim$5~Hz.

A particularly striking feature in the retardance trace is a triple-peak event that spans approximately 2.5~s and comprises a sharp retardance increase, a brief dip, an intermediate peak, a second dip, a recovery, and a sharp decrease. The sarcomere length mirrors this sequence but with an unexpected deviation: while the initial increase and final decrease show the expected anti-correlation with sarcomere length (high retardance corresponding to short sarcomeres), the intermediate peak shows a \textit{positive} correlation, that is, both retardance and sarcomere length increase. This might suggest that the birefringence of the myofibril itself increases during this phase, potentially reflecting a mechanically induced change in molecular alignment (photoelasticity). A detailed biophysical interpretation of these sub-contraction dynamics is beyond the scope of the present work, but the observation demonstrates the potential of high-speed polarimetric imaging to access functional birefringence changes. Such an analysis would be inaccessible to fluorescence imaging.

While in principle fluorescence acquisition rates can be increased, this comes at the cost of higher excitation intensity to maintain a sufficient signal-to-noise ratio. This in turn creates an inherent and unavoidable tradeoff: faster fluorescence imaging accelerates both photobleaching and the risk for phototoxic damage to living cell. In fact, even during our experiments, the fluorescence intensity already decayed by over 20\% within only 15~seconds continuous acquisition (Fig.~\ref{Fig6}b), and even by 70~\% over a 60~s long acquisition (see supplementary Fig.~S7). In contrast, the retardance signal remained constant throughout the entire recording, as the label-free birefringence contrast is inherently unaffected by photobleaching and can be read out at lower light levels, thereby also avoiding phototoxicity. This opens the possibility of arbitrarily long recording sessions, without signal degradation, enabling longitudinal studies of cardiomyocyte maturation, drug response kinetics, and fatigue that are inaccessible to fluorescence-based approaches.

\section{Discussion and Conclusion}
In this work, we presented QP²I, as a dual-mode quantitative phase and polarimetric imaging framework that unites computational reconstruction, polarization optics, and live-cell biology. QP²I is built on a standard DIC microscope, integrating a QWP and a polarization camera. By combining a new total-variation-regularized phase solver with an analytical Stokes-based polarimetric reconstruction, we demonstrated single-shot, video-rate imaging of both isotropic and anisotropic optical properties. The new phase reconstruction addresses the susceptibility of line-integration methods to directional streak artifacts and/or the dependence on reliable segmentation masks. By formulating phase recovery as a single-stage variational problem with a shear-matched finite-difference operator, the method avoids the two-stage integrate-then-correct paradigm of previous approaches~\cite{arnison2000using, yasuhiko2021single} and requires only a single regularization parameter. It can be implemented directly for linear as well as for diagonal shear configurations that are common in commercial DIC microscopes. 

The polarimetric imaging mode extends the same hardware to quantitative birefringence measurements by removing the Nomarski prisms and using a detection QWP to encode the circular polarization component of the sample, which carries the retardance information, into the linearly polarized channels measurable by a conventional polarization camera. This analytical Stokes inversion, computed in a single pass without iterative optimization, recovered both retardance and fast-axis orientation maps at the full camera frame rate. 

The synergy of these two modalities proved particularly powerful for label-free imaging of live cardiomyocyte, where QPI and polarimetry provide complementary information of the same cell. The isotropic phase maps captured overall cell morphology based on height and refractive index differences, which is dominated by the nucleus and was validated against co-registered AFM topography data. The polarimetric maps of the same cells selectively revealed the birefringent contractile myofibrillar architecture that is invisible to conventional phase contrast. Together, these modalities enabled simultaneous monitoring of cellular morphology and sarcomere organization during spontaneous contraction at an effective frame rate of up to 70~fps at full megapixel resolution. Crucially, this frame rate was not limited by biological constraints from the sample, optical constraints of the microscope, or computational constraints of the reconstruction algorithm, but was determined solely by the camera electronics. 

The temporal resolution was sufficient to capture fast contraction dynamics, including double-pump waveforms and complex sub-contraction events that were not resolved in the co-registered fluorescence reference at 10~fps. Moreover, the intrinsic birefringence contrast remained constant throughout extended recordings, while the fluorescence signal decayed by over 70\% within 60~s. These results establish QP²I as a physiologically gentle alternative for sustained contractility analysis.

The sensitivity of QP²I to these detailed dynamics might help answer a long-standing open question about functional muscle biophysics, namely, whether active contraction increases or decreases the optical retardance of the myofibril. Classical bulk measurements have reported both transient increases~\cite{baylor1977optical} and decreases~\cite{chen1989polarization} in birefringence during activation, reflecting competing contributions from myosin head reorientation, filament overlap changes, and fiber geometry~\cite{irving1993birefringence, godfraind1988analysis}. Critically, these studies were confined to bulk preparations or skinned fibers. Our results now suggest that sarcomere length and retardance signal are anti-correlated, however, that sub-contraction events can show transient departures from this anti-correlation (Fig.~\ref{Fig6}). High-speed single-cell polarimetry might therefore allow to revisit the question about contraction induced changes in intrinsic retardance at the individual cardiomyocyte level, disentangling geometric and molecular contributions through simultaneous retardance, orientation, and sarcomere length tracking within single contraction cycles. 

Some limitations of the current implementation merit discussion. The four linear micropolarizers of the camera cannot distinguish left- from right-handed circular polarization, introducing a sign ambiguity in the Stokes component and consequently in the reconstructed orientation of the fast axis $\theta$. In order to maintain quantitative retardance with the same net $\eta$ value for structures at any in-plane orientation (e.g., the same quantitative $\eta$ for all myofibrils at different angles), the polarimetric reconstruction was deliberately designed for $\eta$ without orientation bias, at the cost of reduced $\theta$ determination. However, by rotating the second QWP by 45$^\circ$, the same pipeline could also be used to reconstruct $\theta$ more cleanly at the cost of greater ambiguity in $\eta$. In future works, for samples with low birefringence, this trade-off as well as the sign ambiguity could be completely resolved by using a second synchronized polarization camera with an additional QWP at 45$^\circ$ orientation to directly measure all 4 Stokes parameters (see supplementary material for more detail). 

In the QPI mode, the phase gradient image is extracted via $\operatorname{atan2}$, which wraps at $\pm\pi$ and therefore limits the maximum measurable phase gradient per shear distance. This constraint is relevant for thick, high-contrast samples such as glass microspheres but less so for thin biological specimens. Strongly birefringent samples present an additional subtlety in DIC-to-QPI methods: in DIC microscopy, two orthogonally polarized, laterally sheared beams accumulate different optical path lengths and interfere to produce phase-gradient contrast. However, in birefringent samples, these orthogonally polarized beams also experience different refractive indices at the same spatial position, producing a differential phase delay that is indistinguishable from the isotropic phase gradient. This ambiguity is inherent to any DIC-to-QPI method, and was apparent in our QP²I of single cardiomyocytes, where faint myofibrillar signatures emerged in the phase maps. Importantly, this cross-talk affects only the isotropic phase reconstruction, while the polarimetric imaging mode, without Nomarski prisms, remains unaffected. The availability of both modalities on the same instrument thus provides an intrinsic consistency check: structures appearing in both phase and retardance maps can be identified as birefringent, enabling disambiguation that would not be possible with either modality alone. Another potential limitation is the decrease in effective resolution caused by using a sub-grid of 2$\times$2 polarizer pixels as virtual super-pixel for the reconstruction. Since most experiments throughout this work were carried out with a 100$\times$ magnification, the effective pixel size of the reconstructions was 69~nm which is still well below the optical resolution limit. However, the use of lower magnification objectives might be affected. 

The system operates at 87.7 megapixels per second and is limited only by the camera readout electronics. Polarization-sensitive sensors with higher pixel counts (up to 4096$\times$3000) already exist commercially, and advances in sensor electronics that increase overall pixel throughput per time would directly translate to higher frame rates or larger fields of view of QP²I without any modification to the optical setup or reconstruction software. A natural extension of the current work is volumetric imaging through axial scanning, which would directly enable three-dimensional QPI and polarimetric tomography at rates potentially exceeding those of competing sequential-illumination volume scanning approaches such as tensorial Fourier ptychography~\cite{xu2024tensorial, yeh2024permittivity}, which require hundreds of LED acquisitions per volume. Three-dimensional high-speed polarimetry would be of particular interest for thick muscle tissue. Such volumetric birefringence data could leverage existing three-dimensional muscle morphology analysis tools originally developed for second-harmonic generation (SHG) microscopy~\cite{schneidereit2024cas3d,stein2022software}, which is a significantly more expensive and complex nonlinear imaging modality compared to QP²I.

Beyond muscle cells and tissue, QP²I could open applications of label-free imaging of intrinsic phase and birefringence contrast at video rate, including cancer research (e.g., extracellular matrix organization in tumor microenvironments), neuroscience (e.g., myelination dynamics), and fibrosis (e.g., fibrotic tissue remodeling), where our sustained, non-perturbative observation of molecular-scale organization at video rate could complement or replace current fluorescence-based assays.

\bibliographystyle{unsrt}  
\bibliography{references}

\end{document}